\documentclass[lettersize,journal]{IEEEtran}
\usepackage[dvipsnames, svgnames, x11names]{xcolor}
\usepackage{cite}
\usepackage{amsmath,amssymb,amsfonts}

\usepackage{graphicx}
\usepackage{textcomp}
\usepackage{xcolor}
\usepackage{tabularx}

\usepackage{algorithm}
\usepackage{algpseudocode}

\usepackage{amsmath}

\usepackage{makecell}
\usepackage{subfigure}
\usepackage{multirow}
\usepackage{subfigure}
\usepackage{array}
\usepackage{threeparttable}
\usepackage{cite}
\usepackage{array}
\usepackage[caption=false,font=normalsize,labelfont=sf,textfont=sf]{subfig}
\usepackage{textcomp}
\usepackage{stfloats}
\usepackage{url}
\usepackage{verbatim}
\usepackage{graphicx}

\usepackage{multirow}
\usepackage[dvipsnames]{xcolor}
\usepackage{color}

\usepackage{pifont}
\usepackage{amssymb}
\usepackage{booktabs} 

\def\BibTeX{{\rm B\kern-.05em{\sc i\kern-.025em b}\kern-.08em
    T\kern-.1667em\lower.7ex\hbox{E}\kern-.125emX}}
\usepackage{balance}
\begin{document}
\title{FG-SAT: Efficient Flow Graph for Encrypted Traffic Classification under Environment Shifts}
\author{Susu Cui, Xueying Han, Dongqi Han, Zhiliang Wang,~\IEEEmembership{Member,~IEEE,} \\
Weihang Wang, Yun Li, Bo Jiang, Baoxu Liu and Zhigang Lu
\thanks{This work is supported by Strategic Priority Research Program of Chinese Academy of Sciences (No.XDC02040100), and National Key Research and Development Program of China (2021YFB3101400). This work is supported by the Program of Key Laboratory of Network Assessment Technology, the Chinese Academy of Sciences, Program of Beijing Key Laboratory of Network Security and Protection Technology. \textit{(Corresponding author: Bo Jiang.)}}
\thanks{Susu Cui, Xueying Han, Yun Liu, Bo Jiang, Baoxu Liu and Zhigang Lu are with the Institute of Information Engineering, Chinese Academy of Sciences, Beijing, China, and also with the School of Cyber Security, University of Chinese Academy of Sciences, Beijing, China (e-mail: cuisusu@iie.ac.cn, hanxueying@iie.ac.cn, liyun1996@iie.ac.cn, jiangbo@iie.ac.cn, liubaoxu@iie.ac.cn, luzhigang@iie.ac.cn).}
\thanks{Dongqi Han, Zhiliang Wang are with the Institute for Network Sciences and Cyberspace, BNRist, Tsinghua University, Beijing, China (e-mail: handq19@mails.tsinghua.edu.cn, wzl@cernet.edu.cn).}
\thanks{Weihang Wang is with the University of Southern California, Los Angeles, CA, USA (e-mail: weihangw@usc.edu).}
}

\markboth{Journal of \LaTeX\ Class Files,~Vol.~18, No.~9, September~2020}%
{How to Use the IEEEtran \LaTeX \ Templates}

\maketitle

\begin{abstract}
Encrypted traffic classification plays a critical role in network security and management. Currently, mining deep patterns from side-channel contents and plaintext fields through neural networks is a major solution. However, existing methods have two major limitations: (1) They fail to recognize the critical link between transport layer mechanisms and applications, missing the opportunity to learn internal structure features for accurate traffic classification. (2) They assume network traffic in an unrealistically stable and singular environment, making it difficult to effectively classify real-world traffic under environment shifts. In this paper, we propose FG-SAT, the first end-to-end method for encrypted traffic analysis under environment shifts. We propose a key abstraction, the \textit{Flow Graph}, to represent flow internal relationship structures and rich node attributes, which enables robust and generalized representation. Additionally, to address the problem of inconsistent data distribution under environment shifts, we introduce a novel feature selection algorithm based on Jensen-Shannon divergence (JSD) to select robust node attributes. Finally, we design a classifier, GraphSAT, which integrates GraphSAGE and GAT to deeply learn Flow Graph features, enabling accurate encrypted traffic identification. FG-SAT exhibits both efficient and robust classification performance under environment shifts and outperforms state-of-the-art methods in encrypted attack detection and application classification.

\end{abstract}

\section{Introduction}

Traffic encryption technology plays a significant role in enhancing network security and privacy. By employing encryption protocols, users can have greater confidence in the sensitive transactions and communications they engage in.
To ensure data security, numerous internet services and applications employ various encryption techniques to encrypt transmitted content. However, traffic encryption technology also poses challenges for security managers and internet service providers (ISPs). Firstly, encryption does not guarantee the security of the content itself. Malware can be encrypted and transmitted as easily as legitimate files. In fact, over 80\% of malware spreads through TLS protocol~\cite{oh2021survey}. To bypass firewall and security software detection, many malicious network attacks also employ encryption technology to conceal communication content~\cite{liu2019maldetect}. Therefore, security managers need to identify encrypted traffic in order to inspect malicious encrypted traffic. Secondly, due to the widespread use of smart devices and the emergence of various new application services, network traffic is showing a geometric growth trend~\cite{DBLP:journals/comcom/AbbasiST21}. While ensuring reliable data transmission, ISPs need to identify encrypted traffic in order to provide personalized services to users~\cite{shafiq2020data}, thereby improving network efficiency and enhancing the quality of user services.

As the traffic payload is encrypted, traditional classification methods are no longer effective for encrypted traffic. With the development of machine learning and neural networks, a considerable number of studies show that analyzing the statistical, byte and sequential features of encrypted traffic can also achieve effective classification. According to the feature analysis dimension, we summarize encrypted traffic classification into three methods: (1) Statistics-based methods~\cite{DBLP:conf/icissp/Draper-GilLMG16,DBLP:conf/uss/WangCNJG14,DBLP:conf/uss/HayesD16,DBLP:journals/virology/AndersonPM18,DBLP:conf/ccs/AndersonM16,DBLP:conf/infocom/KorczynskiD14,DBLP:conf/iwqos/ShenWZWL16,DBLP:conf/icnp/ChenZ0ZW19,DBLP:conf/ndss/EdeBCRDLCSP20} extract the side-channel statistical features and header fields to construct machine learning classifiers for traffic classification. 
(2) Byte-based methods~\cite{DBLP:conf/icoin/WangZZYS17,DBLP:conf/isi/WangZWZY17,DBLP:conf/hpcc/CuiJCLLL19,han2023network,DBLP:conf/www/LinXGLSY22} utilize the raw bytes of encrypted traffic and transform the classification task into an image classification task, using neural networks such as convolutional neural network (CNN) or capsule neural network (CapsNet) to learn the spatial features of raw bytes in traffic. 
(3) Sequence-based methods~\cite{DBLP:conf/istel/RamezaniKS20,DBLP:journals/tnsm/ShapiraS21,DBLP:conf/infocom/LiuHXCL19,DBLP:journals/tifs/ShenZZXD21}
treat the traffic as the sequence of packets and  extract the packet length and arrival time as the sequential features, and use the recurrent neural network (RNN) or Encoder-Decoder for classification.

Unfortunately, despite the progress made, there are still two primary limitations in existing works as follows:


\textbf{Overlook the critical link between transport layer mechanisms and applications.}  
Transport layer features are intrinsically linked to applications. To communicate more efficiently, the TCP protocol uses a sliding window and acknowledgment mechanism to send multiple packets at the same time and a single acknowledgment number to confirm receipt of multiple packets. As a result, larger sliding windows are common for streaming applications to complete data transfer quickly, but instant messaging applications typically use a balanced exchange of received and acknowledgment packets. Therefore, encoding structural relationships between packets can be beneficial for encrypted traffic classification. 
However, existing works either do not consider the structural relationships between packets ~\cite{DBLP:journals/cm/RezaeiL19,9896143} or are limited to specific 
classification scenarios and deployment locations~\cite{DBLP:conf/ndss/Fu0023,DBLP:conf/raid/FuLQZZYLD22,DBLP:conf/acsac/PhamHTCT21}, missing the opportunity to learn transport layer features for accurate and generic classification of the encrypted traffic. 

\begin{figure}[!t]
    \centering
    \includegraphics[width=0.9\linewidth]{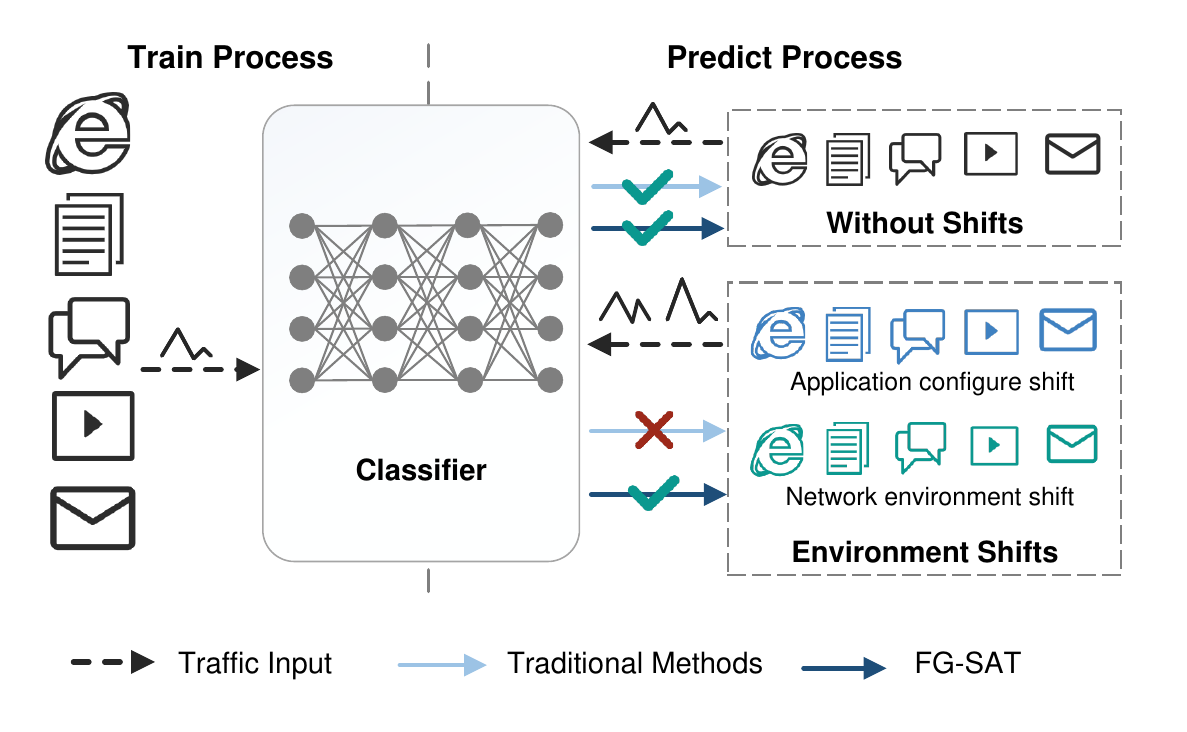}
    \caption{The comparison with the traditional encrypted traffic classification methods. Traditional methods can only identify traffic without shifts that is consistent with the distribution of the training set. In contrast, FG-SAT can identify traffic that does not match the distribution of the training set under environment shifts.}
    \label{partten}
\end{figure}

\textbf{Unrealistic stable and singular network environment.} Current methods mainly focus on traffic classification in a stable and singular network environment, which is unrealistic in representing real-world traffic. As shown in Figure~\ref{partten}, network traffic is susceptible to environmental influences and undergoes long-term dynamic change\cite{9896143,9829791,DBLP:conf/kdd/AndersonM17, DBLP:journals/cn/MalekghainiASLBMMT23}, which we call \textit{environment shifts} in this paper. We define environment shifts as changes in application configuration and/or network environment that cause changes in statistics and feature distribution of traffic within the same class. The application configuration shift includes, but is not limited to, changes in user behaviors (e.g., actions when using a particular application), and the emergence of new software. The network environments shift includes, but is not limited to, changes in protocols and network quality, such as bandwidth. Consider ``browsing'' applications as an example, their environment shifts can include new browser software, different browsing contents, and a change in network bandwidth, compared with the training data environment. If not handled properly, these changes can cause shifts in the distribution of traffic features, leading to poor classification performance. 

To address the abovementioned limitations, we investigate encrypted traffic classification methods with the following challenging goals:
\begin{enumerate}
    \item It is critical to design a flow representation that discovers the intrinsic link between transport layer mechanisms and applications, and learn the relevant features of packets within a flow tied to application.
    \item The flow representation should be i) general enough to represent diverse traffic types, ii) adaptive to feature selection in environment shifts, and iii) independent of encryption protocols.
    \item Since collecting traffic from all possible network environments is impractical, it is necessary to design robust algorithms that can select stable features when the environment shifts to maintain high classification performance. 
\end{enumerate}

In this paper, we propose FG-SAT, the first end-to-end method for encrypted traffic classification in environment shifts. We define a key abstraction, the \textit{Flow Graph}, to characterize the internal relationship structure based on the transport layer mechanisms inside a flow. The Flow Graph provides several key promises: (1) It treats packets as nodes and identifies structural relationships between packets using edges of multiple types that characterize the window and acknowledgment mechanisms. (2) It features a general representation that can represent diverse traffic types, include rich features, and adapt to environment shifts. (3) It also augments node attributes with header fields extracted from the 2-4 layers that are independent of encryption protocols.

To achieve high classification performance in environment shifts, we propose a robust feature selection algorithm to solve the problem of inconsistent data distribution. Specifically, our feature selection algorithm evaluates packet header fields and automatically selects those stable fields as node attributes in the face of environment shifts. We use Jensen-Shannon divergence (JSD), a method for measuring similarity between two probability distributions, to assess the stability of features. Specifically, we compare the distribution differences between inter-class (with environment shifts) and extra-class (with varying traffic types) using JSD. When the extra-class JSD is greater than the inter-class JSD, we consider the features to be stable under environment shifts. Finally, we build a classifier based on graph neural networks (GNN), named GraphSAT, to identify encrypted traffic types. GraphSAT combines GraphSAGE~\cite{DBLP:conf/nips/HamiltonYL17} and GAT~\cite{velivckovic2017graph} to deeply learn the structural relationships and rich node attributes of the Flow Graph, enabling efficient encrypted traffic classification.

\textbf{Our contributions are summarized as follows:}
\begin{itemize}
\item We define a key abstraction, the \textit{Flow Graph}, to represent encrypted flows, which features a general representation for diverse traffic, rich features and encryption protocol independence.
\item We propose a feature selection algorithm, which measures the distribution differences of features by evaluating header fields and selecting stable fields in environment shifts. 
\item We design an encrypted traffic classifier based on GNN, which accurately learns the internal structure and node attributes of the Flow Graph from the raw traffic. 
\item We conduct experiments using publicly available datasets for attack traffic detection and our own collected dataset for application classification. Our method outperforms state-of-the-art methods, increasing accuracy by 6.85\% over pre-training methods and by 15.84\% over traditional deep learning methods. 
\item We evaluate the effects of environment shifts on encrypted traffic classification. The results show that as the environment shifts, all other methods' accuracy decreases, whereas our JSD-based feature selection algorithm increases accuracy by 7.44\%. 
\end{itemize}


\section{Related Work}
In this section, we summarize the related work of encrypted traffic classification and graph-based methods on traffic analysis. Moreover, we compare the classification methods with proposed FG-SAT.
\subsection{Methods on Encrypted Traffic Classification}

\subsubsection{Statistics-based Methods}

Statistics-based methods utilize features like flow duration and packet count for encrypted traffic classification, applying machine learning to distinguish between traffic types. Draper-Gil et al.~\cite{DBLP:conf/icissp/Draper-GilLMG16} use time-related features and the C4.5 algorithm for classifying 12 service types, including VPN traffic. Other works propose features like packet count, size, peak, and time for encrypted web classification~\cite{DBLP:conf/uss/WangCNJG14,DBLP:conf/uss/HayesD16}, and enrich analysis with unencrypted field data for identifying malicious applications~\cite{DBLP:journals/virology/AndersonPM18,DBLP:conf/ccs/AndersonM16} and classifying applications through TLS handshake and certificate information~\cite{DBLP:conf/infocom/KorczynskiD14,DBLP:conf/iwqos/ShenWZWL16,DBLP:conf/icnp/ChenZ0ZW19}. Ede et al.~\cite{DBLP:conf/ndss/EdeBCRDLCSP20} highlight the use of temporal correlations among destination-related features for generating application fingerprints.

Statistics-based methods face limitations including high time latency from processing complete flows, strong feature dependency limiting cross-protocol classification, and a lack of structural feature consideration, reducing accuracy under environment shifts.

\subsubsection{Byte-based Methods}

Byte-based methods input encrypted traffic's raw bytes into neural networks to classify traffic without manual feature extraction, relying on preprocessing and deep learning to uncover traffic byte features. Wang et al.~\cite{DBLP:conf/icoin/WangZZYS17,DBLP:conf/isi/WangZWZY17} utilize the first 784 bytes and CNNs for feature learning in malware and encrypted application classification. Cui et al.~\cite{DBLP:conf/hpcc/CuiJCLLL19} enhance feature learning with CapsNet, surpassing CNNs in spatial and byte feature analysis. Han et al.~\cite{han2023network} use Transformers on n-gram frequency vectors for flow feature learning. Lin et al.~\cite{DBLP:conf/www/LinXGLSY22} introduce ET-BERT, a pre-training model for generic traffic representation learning in a few-shot context.

Byte-based methods simplify feature extraction but face challenges in cross-protocol classification due to plaintext fields in application protocols. These methods may not fully utilize byte information, including timestamps that risk overfitting and hamper generalization across environment shifts. Additionally, while focusing on spatial and temporal traffic byte features, they neglect structural aspects.

\subsubsection{Sequence-based Methods}

Sequence-based methods utilize packet length and time interval sequences, applying models like LSTM and Encoder-Decoder to understand sequence relationships. Ramezani et al.~\cite{DBLP:conf/istel/RamezaniKS20} use the server name from Client Hello packets for fingerprints. Shapira et al.~\cite{DBLP:journals/tnsm/ShapiraS21} develop FlowPic, an image from packet size and arrival times, analyzed with CNN. Liu et al.~\cite{DBLP:conf/infocom/LiuHXCL19} introduce FS-Net, leveraging LSTM-based Encoder-Decoder to explore packet length sequences for classification. However, extracting sequence information from complete flows usually results in high time latency. 

Sequence-based methods analyze packet relationships in a flow but face challenges with environment shifts like network congestion, bandwidth changes, and application updates, affecting sequence consistency and accuracy across environments. Additionally, they sort by packet arrival time, missing varied packet structure representations.

\subsection{Graph-based Methods on Traffic Analysis}

The graph construction methods are closely related to the specific analysis task. In the field of intrusion detection, IPs or domain names are usually used as nodes, and edges are designed based on the communication between nodes to construct a network interaction graph for anomaly detection. However, these methods are not applicable to the general encrypted traffic classification task addressed in this paper. For example, Fu et al.~\cite{DBLP:conf/ndss/Fu0023} implement unsupervised anomaly detection based on network interaction graphs, identifying abnormal edges through a clustering loss function, but it is not suitable for multi-classification tasks. Fu et al.~\cite{DBLP:conf/raid/FuLQZZYLD22} construct a heterogeneous graph, characterizing the communication behavior of hosts and servers through their temporal and spatial features, and then identifying whether the host is infected. However, it has limitations for application classification since a single host can generate various types of application traffic. In application classification, graphs represent communication relationships to classify traffic. Pham et al.~\cite{DBLP:conf/acsac/PhamHTCT21} use IPs and ports as nodes, turning mobile app traffic into a graph for fingerprinting, but this method assumes uniform traffic labels, limiting it to endpoint classification and not suitable for gateway traffic analysis. Shen et al.~\cite{DBLP:journals/tifs/ShenZZXD21} offer a detailed approach, creating flow-specific graphs with packet length and direction, using a GNN to learn flow structures. However, the simplicity of its node features limits effectiveness in complex networks.

In the graph-based traffic analysis, IPs are mainly used as nodes to construct network interaction graphs for analysis. However, these methods has many limitations in general traffic classification tasks, such as classification scenarios and deployment locations. In addition, although Shen et al.~\cite{DBLP:journals/tifs/ShenZZXD21} propose using graphs to represent relationships between packets, it only constructed one burst relationship, and the node features are simplistic, still lacking effective structural relationships and node feature representations.

\begin{figure*}[!t]
    \centering
    \includegraphics[width=\textwidth]{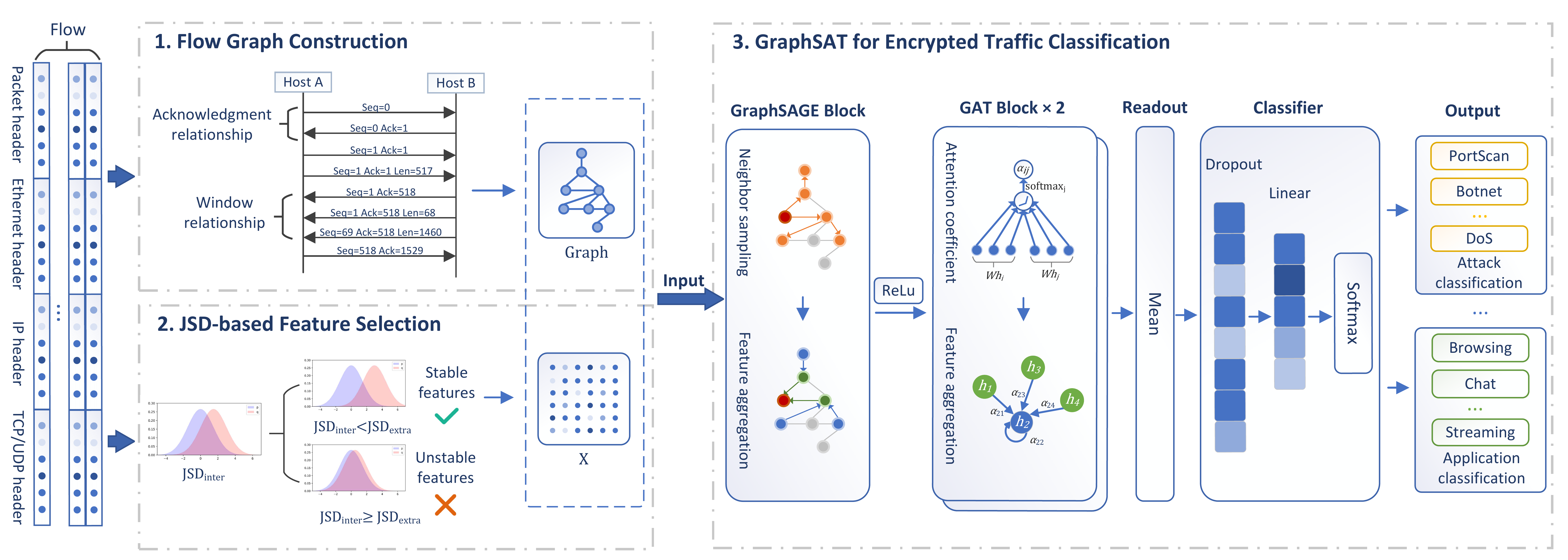}
    \caption{The framework of FG-SAT, including 1. Flow Graph construction, 2. JSD-based feature selection, 3. GraphSAT for encrypted traffic classification.}
    \label{framework}
\end{figure*}

\section{Problem Description and Core Idea}

In this section, we present the problem description of encrypted traffic classification, and then we introduce the framework and the core idea of FG-SAT.

\subsection{Problem Description}

In this paper, we classify encrypted traffic into different types, focusing on scenarios like application and attack type classification. Encrypted traffic consists of packets between clients and servers, encrypted using protocols like TLS, SSH, and Tor, masking application layer content but leaving some header information visible. In this paper, encrypted traffic is composed of multiple flows, where packets of the same flow have the same five-tuple information (source IP, destination IP, source port, destination port, transport layer protocol). We consider flow as the direct object of encrypted traffic classification, and model each flow as a graph for classification using a GNN-based approach, focusing on the structure rather than content. Moreover, we focus on the traffic classification at the firewall and local ISP network nodes, which are typically deployed in enterprise networks. Additionally, our goal is to design an end-to-end system that captures simple traffic information at the edge of the enterprise network for traffic classification. It is worth noting that we only achieve classification through the side-channel information of encrypted traffic without decrypting the traffic. Ultimately, the system can be utilized to optimize resource allocation and perform security monitoring of the enterprise network through encrypted traffic classification.

\subsection{Core Idea}

We introduce FG-SAT to achieve efficient encrypted traffic classification, as shown in Figure~\ref{framework}. Firstly, we define a key abstraction, the \textit{Flow Graph}, that can represents multiple relationships between packets within the same flow and contains rich node attributes. It comprehensively mines the statistical, sequential, and structural features of encrypted traffic. Secondly, we propose a novel JSD-based feature selection algorithm for environment shifts. It can evaluate and select a stable set of features in dynamic and variable traffic environments. Finally, we establish a GNN-based encrypted traffic classifier, named GraphSAT, which fully explores the internal relationships between different nodes within the Flow Graph and rich node attributes, achieving efficient encrypted traffic classification.

During Flow Graph construction, each flow is viewed as a graph, capturing its statistical, sequential, and structural features for a comprehensive feature representation. Packets serve as nodes, with relationships defined by transport layer mechanisms and node attributes derived from header fields like packet length, arrival time, direction, TTL, and window size. These attributes are essential for encrypted traffic classification~\cite{cui2022only}. The transport layer's window size, vital for classification, indicates the communication type, with the TCP protocol's sliding window mechanism adjusting for efficient, reliable transmission tailored to the application's needs, such as larger windows for streaming traffic to ensure high throughput. To depict the flow's internal structure, we establish two types of relationships:
\begin{itemize}
    \item \textbf{Window relationship:} The client or server sends multiple packets continuously, and these packets are connected in sequence based on their arrival time to form the window relationship.
    \item \textbf{Acknowledgment relationship:} The client or server acknowledges the received packets so that the packet and its corresponding acknowledgment packet form the acknowledgment relationship.
\end{itemize}

Due to the environment shifts, including changes in application configure and network environment, network traffic is in a state of constant change, causing traditional encrypted traffic classification methods to lose accuracy~\cite{DBLP:conf/kdd/AndersonM17,DBLP:journals/cn/MalekghainiASLBMMT23}. To address this, we introduce a JSD-based feature selection algorithm that assesses feature stability across different environments by measuring inter-class and extra-class distribution differences with JSD. This algorithm helps select stable distribution features as final node attributes amidst environment shifts.

We also develop GraphSAT, a GNN-based classifier, combining GraphSAGE and GAT to analyze Flow Graph's structure and node attributes. GraphSAGE samples neighbors to enhance generalization and reduce memory use, while GAT assesses neighbor importance. GraphSAT effectively and accurately classifies encrypted traffic types.

\section{Construction and Analysis of Flow Graph}

In this section, we provide a detailed description of the process of Flow Graph construction. Additionally, we illustrate the differences between Flow Graphs among various types of traffic through analysis of specific examples.

\subsection{Flow Graph Construction}

\begin{figure}[!t]
    \centering
    \includegraphics[width=\linewidth]{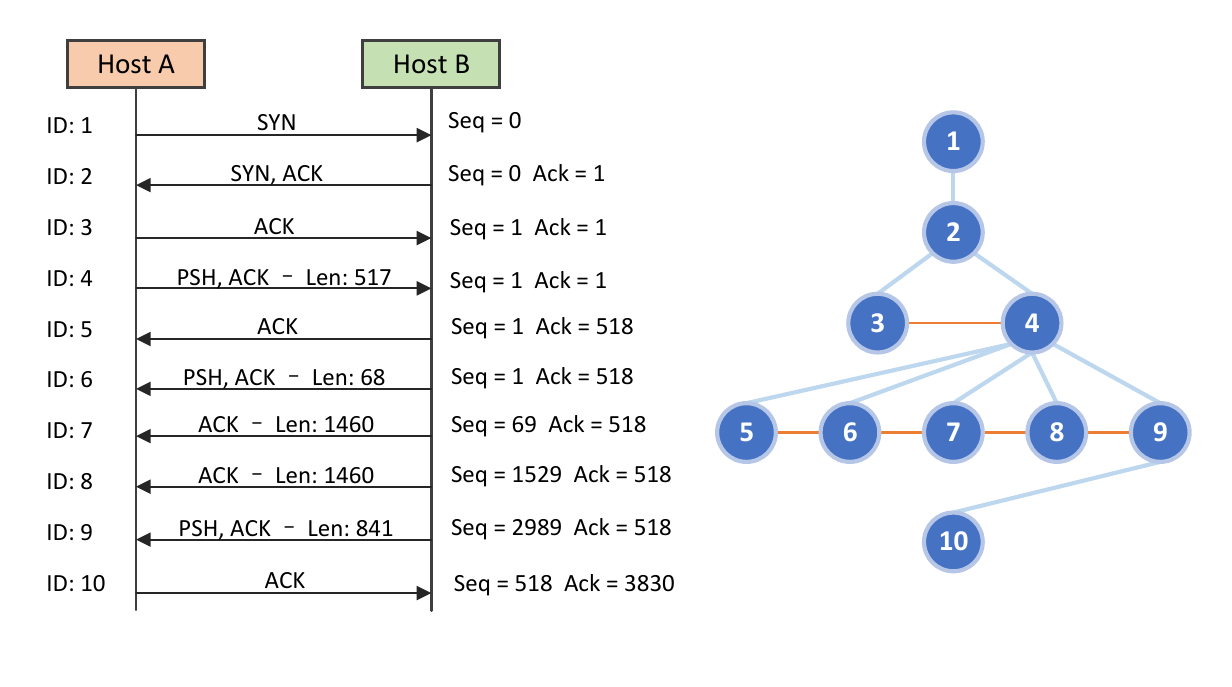}
    \caption{An example of Flow Graph on TCP flow.}
    \label{TCP}
\end{figure}

\begin{figure}[!t]
    \centering
    \includegraphics[width=0.85\linewidth]{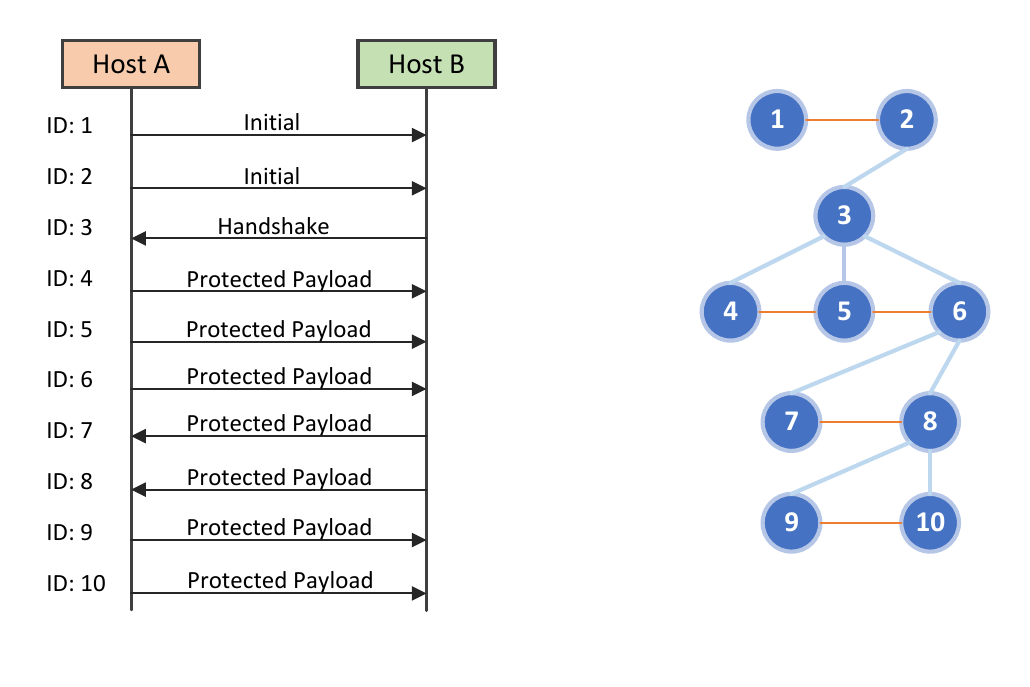}
    \caption{An example of Flow Graph on UDP flow.}
    \label{UDP}
\end{figure}
To achieve an end-to-end encrypted traffic classification, we convert the raw traffic into the Flow Graph. Firstly, we aggregate traffic based on flow granularity. In long-term services such as file transfers or malicious C\&C communication, the duration of a flow can be as long as several hours. In order to perform efficient traffic classification, we only extract the first $n$ packets of a flow to establish a Flow Graph, where $n$ is considered as a hyperparameter of our method.

After traffic aggregation, we construct the Flow Graph consisting of no more than $n$ packets. In this paper, a Flow Graph is defined as $G(V, E)$ consisting of $n$ nodes and $m$ edges. $V=\{v_{1}, v_{2}, … , v_{n}\}$ is the set of nodes, with each node representing a packet within the flow. The node attributes of the node $v$, denoted as $x_{v}$, is a vector representation of the header fields, including packet header, ethernet header, IP header and TCP/UDP header. However, some fields in the packet header may not be effective in environment shifts. Therefore, we use the JSD-based feature selection algorithm to evaluate all the fields and select the stable fields as the final node attributes. More details on the JSD-based feature selection algorithm are presented in Section~\ref{sec:jsd}.

$E=\{e_{1}, e_{2}, …, e_{m}\}$ is the set of edges, where each $e \in E$ represents the relationship between packets, including window relationship and acknowledgment relationship, defined as follows:
\begin{itemize}
    \item \textbf{Window relationship:} When multiple packets are continuously sent from the sender to the receiver, they are connected in sequence according to the arrival time and form a window relationship. All packets in a single window are contiguous and in the same direction. In addition, in TCP packets, they also have the same ACK number.
    \item \textbf{Acknowledgment relationship:} When the receiver confirms and responds to the received packets, the packets and their corresponding acknowledgment packet form a acknowledgment relationship. They are in opposite directions. In TCP packets, the acknowledgment relationship can be associated based on the SEQ and ACK number. UDP packets consider adjacent packets, which are sent in opposite directions, as having an acknowledgment relationship.
\end{itemize}

\begin{figure*}[!t]
    \centering
    \includegraphics[width=0.85\textwidth]{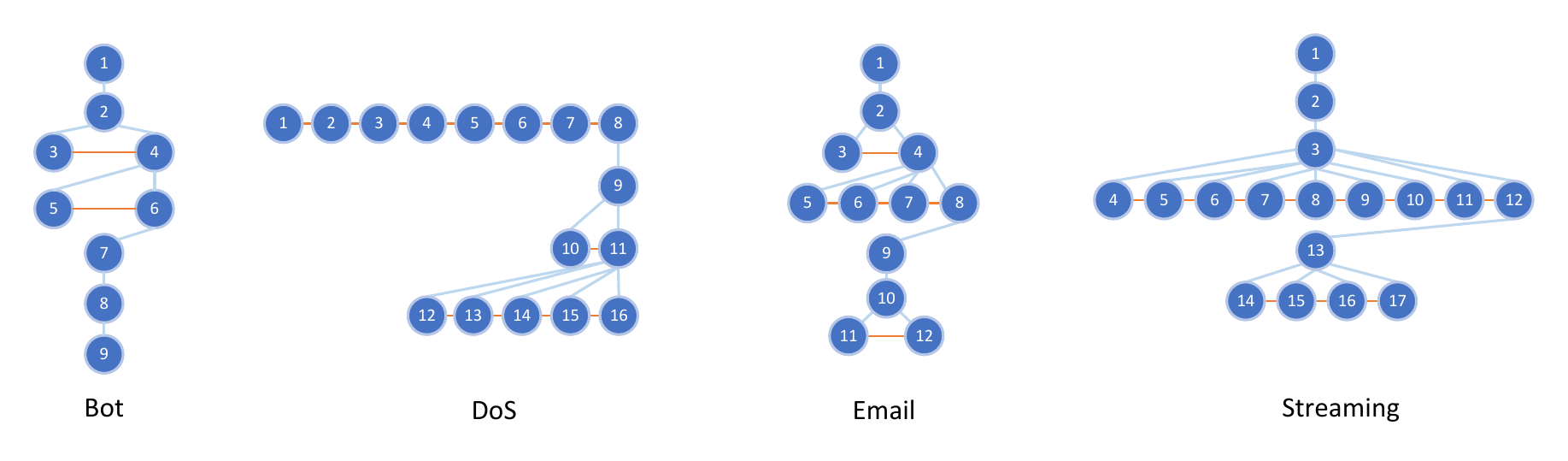}
    \caption{Examples of Flow Graph with different labels.}
    \label{example}
\end{figure*}

We analyze TCP and UDP flows to understand their transmission behaviors and Flow Graph construction. For TCP flow, depicted in Figure~\ref{TCP}, The packets inside the flow are the nodes of the graph, each represented by header field patterns. Packets in the same window share direction, continuity, and ACK number, forming window relationships based on arrival times. Acknowledgment relationships are built when a receiver's ACK number matches the sender's SEQ number plus TCP packet length. Hence, nodes within the same window, such as nodes 5-9, share both window and acknowledgment relationships, typically with a common node.

For UDP flow, shown in Figure~\ref{UDP}, the absence of SEQ and ACK fields means reliability mechanisms of TCP are not present. Here, packets that share direction and continuity are considered in the same window, and adjacent packets in opposite directions have acknowledgment relationships. As with TCP, nodes in window relationships share acknowledgment relationships with a specific node, simplifying the construction of UDP Flow Graphs.

\subsection{Analysis on Flow Graph}

In this section, we describe and compare the Flow Graphs of different encrypted traffic types. We focus on the overall graph structure rather than node attributes. Based on the definition of Flow Graph mentioned above, we construct Flow Graphs for four types of encrypted traffic, Bot, DoS, Email, and Streaming, as shown in Figure~\ref{example}. It can be seen that different types of encrypted traffic have different graph structures. Bot traffic is primarily controlled by streamlined commands, so it has a small window size and frequent acknowledgment. DoS sends packets with large window size at the beginning of the TCP connection. Emails generally transmit plaintext content, thus, the upload and download windows are similar in size. Streaming servers typically send large window size packets to clients because they download data more than they upload. Thus, different types of traffic exhibit different internal transmission structures based on the content of their upper-layer applications. Flow Graphs can clearly represent the internal structure of traffic by proposed edge relationships.

\section{JSD-based Algorithm for Feature Selection}
\label{sec:jsd}

\begin{algorithm}[!htb]
\caption{JSD-based Feature Selection Algorithm}
\label{jsd_feature_selection}
\begin{algorithmic}[1]
\State \textbf{Input:} Training set data $T$, label $L$, candidate feature set $F$, environment label $EL$ (optional), number of samples for each class $Len$.
\State \textbf{Output:} Stable feature set under environment shifts $Z$.
\State \textbf{Variables:} 
\State \hspace{4mm} $T_{I}, T_{II}, T_{III} \gets$ subsets of $T$, all three mutually exclusive.
\State \hspace{4mm} $EL_{I}, EL_{II} \gets$: subsets of $EL$, two are mutually exclusive.
\State \hspace{4mm} $FD \gets$: JSD difference matrix, $FD[f, l]$ denotes the JSD difference of label $l \in L$ in feature $f \in F$.

\Procedure{JSD difference calculation}{$T$, $L$, $EL$, $F$}
\State $EL_{I}, EL_{II} \gets$ two empty lists.
\If{$EL$ is True}
\State Randomly divide $EL$ into $EL_{I}$ and $EL_{II}$, which of them have different environment settings.
\EndIf
\For{$l$ in $L$}
\State $T_{I} \gets$ data labeled as $l$ and $el$ belongs to $EL_{I}$.
\State $T_{II} \gets$ data whose label is $l$ and $el$ belongs to $EL_{II}$.
\State $TIII \gets$ data whose label is not $l$.
\For{$f$ in $F$}
\State $FD[f][l] \gets$ $JSD(T_{I}[f], T_{II}[f]) - $
\Statex \hspace{33mm} $JSD(T_{I}[f], T_{III}[f])$
\EndFor
\EndFor
\State \textbf{return} $FD$.
\EndProcedure

\Procedure{Stable Feature Selection}{$FD$, $Len$}
\For{$fd$ in $FD$}
\State $diff \gets$ $\sum_{i=1}^{n}{fd_i \times Len_i}$, where $n$ is the number of types.
\If{$diff < 0$}
\State Add the feature $f$ to $Z$.
\EndIf
\EndFor
\State \textbf{return} $Z$.
\EndProcedure
\end{algorithmic}
\end{algorithm}

In this section, we introduce a feature selection algorithm based on JSD. It aims to evaluate the inter-class distribution differences and extra-class distribution differences of features in environment shifts, and selects stable features to provide accurate and stable feature representation. JSD is a method for measuring the similarity between two probability distributions. It is a variation of the KL divergence that solves the problem of KL divergence's asymmetry~\cite{DBLP:journals/entropy/Nielsen20}. Assuming there are two probability distributions, $P$ and $Q$, with probability density functions of $p(x)$ and $q(x)$ at point $x$, respectively. The definition of JSD is defined as follows:
\begin{align}
JSD(P,Q) &= \frac{1}{2}(D_{KL}(P||M) + D_{KL}(Q||M)) \\
D_{KL}(P,Q) &= \sum_{i} P(i) \log \frac{P(i)}{Q(i)}\\
M &= \frac{1}{2}(P+Q)
\label{JSD}
\end{align}
where $D_{KL}$ denotes the KL divergence, and $M$ is the intermediate distribution between $P$ and $Q$. The value of JSD ranges from 0 to 1, with 0 indicating that the two distributions are identical and 1 signifying that both distributions are completely different. The greater the similarity between two probability distributions, the closer the value of JSD is to 0.

The JSD-based feature selection algorithm is described in Algorithm~\ref{jsd_feature_selection} for multi-class classification tasks. It is important to note that we only use labeled data from the training set for feature evaluation and selection. We evaluate the inter-class and extra-class JSD of each feature for every class. We create three datasets from the training set, where $T_{I}$ and $T_{II}$ are used to evaluate the inter-class JSD of the traffic and $T_{I}$ and $T_{III}$ are used to evaluate the extra-class JSD. For example, when we evaluate the difference in feature distribution between ``browsing'' and other 
labels, we can divide ``browsing'' traffic into two non-overlapping datasets $T_{I}$ and $T_{II}$ based on environment shifts. Specifically, we randomly split the factors of ``browsing'' according to the environment shift factors, as shown in Table~\ref{dataset}, into two non-overlapping sets: $EL_{I}$ and $EL_{II}$. Subsequently, we allocate all the traffic belonging to $EL_{I}$ into $T_{I}$ and all the traffic associated belonging to $EL_{II}$ into $T_{II}$.
If there is no environment shifts label, $T_{I}$ and $T_{II}$ are obtained by randomly dividing ``browsing'' traffic. In contrast, $T_{III}$ is the traffic of other labels traffic, in this example, other labels include chat, emial file and streaming.

We then calculate the inter-class and extra-class JSD for each feature using $T_{I}$, $T_{II}$, and $T_{III}$, obtaining the distribution stability of each feature in varying environments. If the JSD of the inter-class distributions is smaller than the JSD of extra-class ones, the feature is considered stable in varying environments and is retained for subsequent traffic classification. To perform multi-class encrypted traffic classification, we also need to evaluate the comprehensive stability of features across multiple classes, weighted according to the number of samples for each class. 

In this paper, we use header fields as node attributes, noting their distribution can vary across different environments. For instance, the type of content in instant messaging, like text or multimedia, impacts packet length distribution, and network bandwidth speed influences arrival time fields. To address this, we apply a feature selection algorithm described in Algorithm~\ref{jsd_feature_selection} for encrypted traffic classification tasks, choosing stable header fields as final node attributes to accurately characterize encrypted traffic.

\section{GraphSAT Classifier for Encrypted Traffic Classification}

GraphSAT takes the Flow Graph as input and learns the internal structure features and rich node attributes to achieve efficient and accurate encrypted traffic classification. The structure of GraphSAT mainly consists of GraphSAGE Block, GAT Block, Readout, and Classifier.

\subsection{GraphSAGE Block}
We use GraphSAGE to collect local information on nodes in the Flow Graph and calculate node embeddings. GraphSAGE is a technique used to learn a node representation method that involves sampling and aggregating features from its local neighbors. The GraphSAGE block comprises two processes, neighbor sampling, and feature aggregation.

\textbf{Neighbor sampling:} It samples a fixed number of neighbors $S$, for each node to ensure efficient computation, regardless of the node's degree. For nodes with fewer than $S$ neighbors, sampling with replacement (SWR)~\cite{razavi2019imputation} ensures $S$ nodes are sampled; for those with more, sampling without replacement (SWOR)~\cite{zmigrod2021efficient} is applied. GraphSAGE uses a constant $K$ to define the hop number for neighbor sampling, enhancing distant neighbor information capture. We set $K=2$, sampling $S_{1}$ first-order and $S_{2}$ second-order neighbors for each target node.

\textbf{Feature aggregation:} It creates the target node's embedding by averaging the feature vectors of sampled neighbors, moving from second-order to first-order neighbors before reaching the target node. Average aggregation is employed to combine neighbor embeddings dimension-wise, followed by a non-linear transformation. The definition is given as follows:
\begin{equation}
h_{v}^{k}=\sigma(W\cdot \text{MEAN}(h_{v}^{k-1}\cup h_{u}^{k-1}), {\forall}u\in\mathcal{N}_{v})
\end{equation}
where $h_{v}^{k}$ is the $k$-th layer node $v$ feature vector, $\mathcal{N}_{v}$ is $v$'s neighbor set, $\sigma$ denotes the activation function, and $W$ is the $k$-th layer's trainable weight matrix.

\subsection{GAT Block}

We obtain the first layer of node embeddings using GraphSAGE. Next, we further calculate the importance of nodes and generate new node embeddings using GAT. It assigns different learning weights to different neighbors for learning the interrelationships between nodes. GAT Block consists of two processes, calculating attention coefficients and feature aggregation.

\textbf{Attention coefficient:} Each attention coefficient is learned through the self-attention mechanism, where each node in the graph learns the weight for each of its neighbors based on their respective feature vectors. The definition of the attention coefficient is as follows:
\begin{equation}
\alpha_{i,j} = \frac{\exp\left(\mathrm{LeakyReLU}\left(a[Wh_i || Whj] \right)\right)}{\sum{k \in \mathcal{N}_i} \exp\left(\mathrm{LeakyReLU}\left(a^T [Wh_i || Wh_k] \right)\right)}
\end{equation}
where $\alpha_{i,j}$ represents the attention coefficient for the edge between nodes $i$ and $j$, $W$ is a weight matrix, $h_i$ and $h_j$ are the feature vectors for nodes $i$ and $j$, respectively, $\mathcal{N}_i$ is the set of neighbors of node $i$, $||$ represents concatenation, and $\mathrm{LeakyReLU}$ is the leaky rectified linear unit activation function. The vector $a$ is a learnable parameter vector that is shared across all nodes and is used to calculate the compatibility score between node attributes.

\textbf{Feature aggregation:} According to the attention coefficient, we aggregate the node attributes by weighting and summing them to obtain the embedded representation of the aggregated node. It is defined as follows:
\begin{equation}
h_i^{'} = \sigma\Big(\sum_{j\in N_i} \alpha_{ij} W h_j \Big)
\end{equation}
where $h_{i}'$ is the new feature output by GAT for each node $i$, which incorporates neighbors information, and $\sigma$ is the activation function. In addition, we use multi-head attention to enhance the node attributes, which is defined as follows:
\begin{equation}
\overrightarrow{h_i} = \big\Vert_{k=1}^K h_i^{(k)}
\end{equation}
where $K$ is the number of attention heads. The final output $\overrightarrow{h_i}$ is the concatenation of the outputs from all attention heads.

\subsection{Readout}
After the GraphSAGE and GAT processing, we generate node embeddings for the Flow Graph. To represent the whole graph, we use a global mean pool to derive the graph's overall representation vector, defined as:
\begin{equation}
\mathcal{R}(\mathbf{H}) = \sigma\big(\frac{1}{N} \sum_{i=1}^{N} \mathbf{h}_i\big)
\end{equation}
where $\mathcal{R}(\mathbf{H})$ is the whole graph's embedding, $\mathbf{h}_i$ represents the i-th node's feature, and $\sigma$ refers to the sigmoid function.

\subsection{Classifier}

The classifier consists of dropout, linear, and softmax layers to prevent overfitting and classify encrypted traffic. Dropout layers randomly omit nodes to reduce interactions and overfitting, serving as regularization. Linear layers map high-dimensional data to a lower-dimensional label space, with the softmax layer finalizing classification. We concentrate on classifying encrypted traffic by application type (such as browsing, email) and attack type (such as DoS, PortScan).

\section{Experiment and Evaluation}

In this section, we evaluate and compare the proposed FG-SAT. Firstly, we describe the experiment setting. Next, we perform a comparison with baseline methods, analysis on environment shifts and the JSD-based feature selection algorithm, analysis on GraphSAT, analysis on open-world classification and few-shot analysis to fully evaluate the FG-SAT.

\subsection{Experiment Setting}

\subsubsection{Datasets}

\begin{table*}[!htb]
\centering
\caption{The statistics of experimental datasets.}
\begin{tabular}{c|c|c|c|c|c|c}
\hline
\multicolumn{3}{c|}{Attack Detection}&\multicolumn{4}{c}{Application Classification}\\
\hline
Label&Count&Ratio&Label&Factor&Count&Ratio\\
\hline
\multirow{2}*{Benign}	&\multirow{2}*{50000}	&\multirow{2}*{15.86\%}	&\multirow{2}*{Browsing}	&\textbf{Content}: [Blog, Map, Picture, Video],  \textbf{Bandwidth}: [20Mbps, 100Mbps]	&\multirow{2}*{12272}	&\multirow{2}*{66.88\%}\\
&&&&\textbf{Browser}: [Google, Edge], \textbf{Speed of opening new pages}: [1s,  10s]&&\\
\hline
\multirow{1}*{DDoS}	&\multirow{1}*{50000}	&\multirow{1}*{15.86\%}	&\multirow{1}*{Chat}	&\textbf{Content}: [Picture, Video, Text, Voice], \textbf{APP}: [Facebook, WeChat]&\multirow{1}*{1518}	& \multirow{1}*{8.27\%}\\
\hline
\multirow{1}*{DoS}	&\multirow{1}*{50000}	&\multirow{1}*{15.86\%}	&\multirow{1}*{Email}	&\textbf{APP}: [Outlook, Google], \textbf{Action}: [Send, Read]&\multirow{1}*{4810}	&\multirow{1}*{13.11\%}\\
\hline
\multirow{1}*{Spoofing}&	\multirow{1}*{50000} &	\multirow{1}*{15.86\%} &	\multirow{1}*{File} &	\textbf{Content}: [MP4, Word, Zip file], \textbf{Action}: [Upload, Download]&\multirow{1}*{1179}&	\multirow{1}*{6.43\%}\\
\hline
\multirow{1}*{Recon}&	\multirow{1}*{50000}&	\multirow{1}*{15.86\%}	&\multirow{1}*{Streaming}& \textbf{Resolution}: [270, 480, 720, 1080], \textbf{Playback}: [0.5, 1.0, 1.25, 1.5, 2.0] &\multirow{1}*{1686}&	\multirow{1}*{5.31\%}\\
\hline
Mirai&	50000&	15.86\%	&\multicolumn{4}{c}{}\\
\hline
Web attack&	15239&	4.83\%	&\multicolumn{4}{c}{}\\
\hline
\hline
Overall&	315239&	100\%		&Overall&&	18349&	100\%\\
\hline
\end{tabular}
\label{dataset}
\end{table*}

In this paper, we evaluate our proposed method through three encrypted traffic classification tasks: \textit{encrypted traffic attack detection}, \textit{encrypted traffic application classification}, and \textit{encrypted traffic application classification under environment shifts}. 

For the encrypted traffic attack detection, we use the publicly available CIC-IOT2023~\cite{neto2023ciciot2023} dataset, which contains benign and the most up-to-date common attacks resembling real-world traffic. To construct environment shifts in attack detection, we partition data according to the IP pair method, ensuring that traffic from the same IP pair exists only in either the training set or the test set. Additionally, we enhance the realism of environment shifts by supplementing the ``Benign'' class traffic with our own collected dataset (APP-SHIFTS dataset), which includes various environment shift factors.

To evaluate the environment shifts, we newly collect an application dataset, named APP-SHIFTS dataset, for encrypted traffic application classification and encrypted traffic application classification under environment shifts. We capture multiple encrypted application traffic from a campus network environment. Specifically, when the user accesses an encrypted application through the PC, the generated traffic is saved to the data server by the traffic capture tool. In addition, we construct environment shifts by setting different factors for each type of traffic, such as content, bandwidth, and resolution, totaling 72 types of environment factors. The statistics of the datasets are shown in Table~\ref{dataset}. 

\subsubsection{Baseline Methods}
In order to evaluate and compare the performance of FG-SAT, we summarize the following baseline methods:

\begin{itemize}
    \item \textbf{FlowPrint:} It is a semi-supervised approach for mobile application fingerprinting using temporal correlations to generate application fingerprints.
    \item \textbf{1dCNN:} It extracts the first 784 bytes of encrypted flows, converting them into a picture and using 1dCNN to learn spatial features for classification.
    \item \textbf{CapsNet:} It utilizes CapsNet to learn spatial and location features of encrypted flows for classification.
    \item \textbf{GTID:} It computes n-gram frequency of packet payloads, converting them into vectors and applying transformers for feature learning.
    \item \textbf{ET-BERT:} It is a pre-training model that learns generic traffic representations from unlabeled traffic and fine-tunes for specific tasks.
    \item \textbf{FlowPic:} It generates an image for each flow from packet size and arrival time, using CNN for classification.
    \item \textbf{FS-Net:} It extracts packet length sequences and employs an LSTM-based Encoder-Decoder for classification.
    \item \textbf{Rosetta:} It applies TCP-aware augmentation and self-supervised learning to enhance TLS traffic classification in diverse networks.
    \item \textbf{GraphDApp:} It is a GNN-based method using packet length and direction to learn structural packet relationships for DAPP identification.
\end{itemize}

\subsubsection{Evaluation Metrics and Implementation Details}

We assess FG-SAT's performance using Accuracy (Acc), Precision (Pre), Recall (Rec), and $F_{1}$, employing Macro Average to mitigate bias from class imbalances. Prediction speed is gauged by the time to predict 100 flows, and model complexity and memory usage are evaluated through trainable parameters. Our experiments rely on 5-fold cross validation for robustness.

For Flow Graph construction, we limit flows to 20 packets. In GraphSAT, hidden layers for both GraphSAGE and GAT are set at 128. We use a batch size of 128, a learning rate of 0.003, and cap epochs at 100, incorporating a 0.7 dropout ratio. Parameter fine-tuning details are in section~\ref{sec:param}.

\subsection{Comparison with Baseline Methods}

\begin{table*}[!htb]
\setlength\tabcolsep{4pt}
\centering
\caption{The results and comparison with baseline methods.}
\begin{threeparttable}

\begin{tabular}{c|c|c|c|c|c|c|c|c|c|c|c|c|c|c|c}
\hline
\multirow{2}*{Categories} & \multirow{2}*{Method} & \multicolumn{4}{c|}{Attack Detection} &\multicolumn{4}{c|}{APP Classification}&\multicolumn{6}{c}{APP Classification under Environment Shifts}\\
\cline{3-16}
&&Pre&Rec&$F_{1}$ &Acc&Pre&Rec&$F_{1}$ &Acc&Pre&Rec&$F_{1}$ &Acc&Time (ms)&Params\\
\hline
Statistics-based&FlowPrint	&0.4896	&0.4548	&0.4421	&0.4458	
&0.8125  &  0.8180 &   0.8082 	&0.8867
&0.7211&0.4093&0.4720&0.7370	&594.36&-\\
\hline
\multirow{4}*{Byte-based}&1dCNN&0.6095&0.5427&0.5440&0.5515&0.6728&0.7774&0.7105&0.8017&0.4781&0.5051&0.4913&0.7781&5.16&5825413\\
&CapsNet&0.6117&0.5605&0.5716&0.6126 &0.7643&0.7961&0.7791&0.8460 &0.7844&0.6998&0.7251&0.8160&185.57&7592976\\
&GTID	&0.6489	&0.6106	&0.6229	&0.6127	
&0.7836	&0.8668	&0.8088	&0.8711	
&0.8202	&0.7503	&0.7728	&0.8220	&289.74	&893797\\
&ET-BERT&	0.7759&	0.7766&	0.7607	&0.7823
&	0.9017	&0.8834	&0.8915&	0.9214
&	0.8180&	0.7928&	0.8022&	0.8614&	396.23	&132129797\\
\hline
\multirow{3}*{Sequence-based}&FlowPic		&0.7058	&0.6352	&0.6515	&0.6471	&0.7749	&0.7944	&0.7828	&0.8695 
&0.8296	&0.6259	&0.7022	&0.8243	&9.63	&1597219\\
&FS-Net		
&0.7182	&0.6409	&0.6582&	0.6528
&0.8106	&0.7313	&0.7513&	0.8597
&0.7482&	0.6972&	0.7115&	0.8377	&210.27	&2250451\\
&Rosetta	
&0.6606& 0.6045& 0.6197& 0.6160
&0.8746 &   0.7558  &  0.8033&	0.8664
&0.6409&	0.5392&	0.5587&	0.7544	&131.97& 11402479\\
\hline
\multirow{3}*{Graph-based}&GraphDApp&0.7457&    0.6783&    0.6922&0.6924& 0.8177&    0.8775&    0.8408&0.8618& 0.6766&    0.6026&    0.5916&0.8154&9.42&\textbf{35205}\\

&FG-SAT full&	0.7665&	0.7705&	0.7556	&0.7764
	&0.8916	&0.8897&	0.8906	&0.9468
	&	0.7616&	0.7210	&0.7380	&0.8406	&5.03&	44293\\
 
&\textbf{FG-SAT}	&\textbf{0.8357}&	\textbf{0.8427}	&\textbf{0.8330}	&\textbf{0.8508}	
&\textbf{0.9028}	&\textbf{0.8956}	&\textbf{0.8991}	&\textbf{0.9516}	
&\textbf{0.8250}	&\textbf{0.8049}&	\textbf{0.8124}&	\textbf{0.8979}	&\textbf{4.82}&	43013\\

\hline
\end{tabular}
\label{baseline}
 \begin{tablenotes}
        \footnotesize
        \item * \textbf{Bold} denotes the best reuslt.
      \end{tablenotes}
\end{threeparttable}
\end{table*}

We evaluate FG-SAT and baseline methods in attack detection, application classification, and application classification under environment shifts, focusing on accuracy, classification speed, and model parameters. Moreover, we compare \textit{FG-SAT}, utilizing JSD-based feature selection, with \textit{FG-SAT full} without feature selection. Accuracy reflects classification capability, while time indicates efficiency critical for encrypted traffic analysis. Model parameters gauge complexity and memory consumption, highlighting the importance of lightweight classifiers. Results are detailed in Table~\ref{baseline}.

In terms of accuracy, we evaluate the methods using macro average of four metrics, Pre, Rec, $F_{1}$, and Acc, which are cross-validated to average the results. We observe that FG-SAT shows the best performance in all encrypted traffic classification tasks, reaching an accuracy of 0.8508 in the attack classification, 6.85\% higher than the best baseline method. The accuracy of 0.9516 in application classification is 3.02\% higher than the best baseline method, and the accuracy of 0.8979 in the application classification under environment shifts is 3.65\% higher than the best baseline method. However, the baseline method with the best results, ET-BERT, is a pre-training model for traffic representation learning, and its high-performance results from a large amount of traffic collected. Except for the pre-training method ET-BERT, FG-SAT is significantly better than all other traditional baseline methods.

In terms of time, we investigate the time it takes to predict a flow. FG-SAT takes only 4.82ms to predict 100 flows, which is the shortest time they take. It indicates that FG-SAT can simultaneously classify encrypted traffic quickly.

In terms of parameters, we count the number of trainable parameters during the model run, and the number of parameters in FG-SAT is only second to GraphDAPP, which has just one node attribute. Moreover, the number of parameters in FG-SAT is 3/10,000 of that in ET-BERT. Thus, FG-SAT can achieve the lightweight classification of encrypted traffic, which greatly reduces the memory requirements of devices in practical deployments.

Furthermore, comparing FG-SAT with FG-SAT full, we observe that after JSD-based feature selection, FG-SAT is able to achieve better performance capabilities. In particular, the accuracy of FG-SAT is 7.44\% higher than that of FG-SAT full under environment shifts. Thus, the JSD-based feature selection algorithm is able to select robust features and improve the generalization ability of the method.

Overall, FG-SAT achieves the best performance in terms of combined accuracy, time and parameters.
\begin{figure*}[!t]
    \centering
    \includegraphics[width=0.87\textwidth]{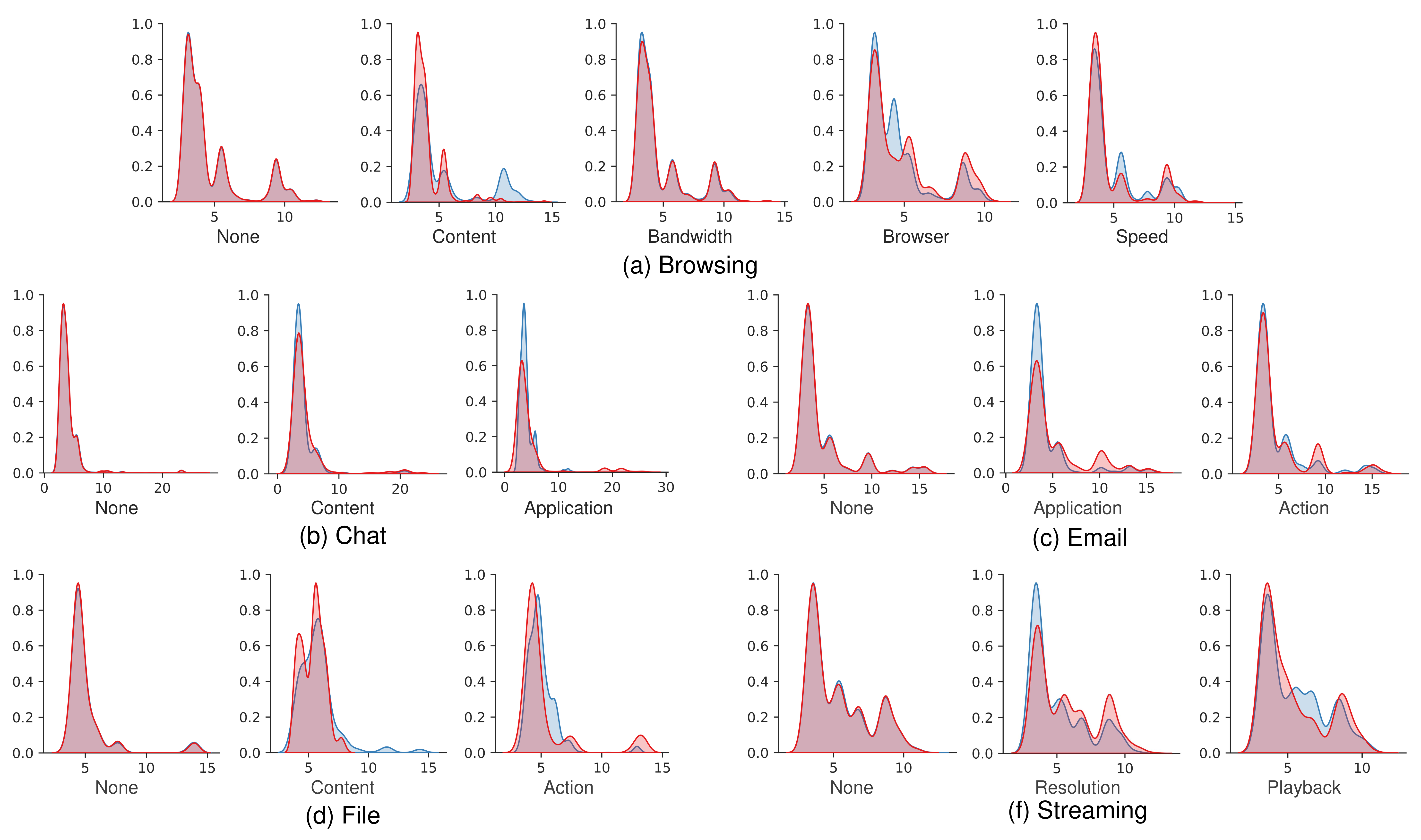}
    \caption{Differences in data distribution under various environment shifts.}
    \label{diff}
\end{figure*}

\subsection{Analysis on Environment Shifts and JSD-based Algorithm}

In this section, we first analyze the impact of environment shifts on the data distribution, and then perform the analysis of the JSD-based feature selection algorithm for a specific ``browsing'' traffic identification scenario. Finally, we compare the JSD-based algorithm with existing feature selection algorithms to evaluate the performance.

\subsubsection{Analysis on Environment Shifts}

In the collected encrypted traffic application classification dataset, environment shifts are constructed by altering various factors. In this section, we analyze the impact of these factors on data distribution. Different shift settings are implemented for five types of encrypted traffic, such as bandwidth, application, and action. Header fields are selected as node attributes in this paper, therefore, the distribution status of the header fields of packets is compared to clearly reflect data distribution differences. Since header fields represent multi-dimensional features, data distribution is measured by comparing the distance of packets to the cluster centroid through clustering methods.
Note that we only use clustering to obtain the centroid for calculating data distribution, rather than employing clustering for classification. Therefore, we set the number of clusters to 1.
For instance, under the content shift of ``browsing'' traffic, K-means~\cite{ahmed2020k} clustering is performed on the traffic of two groups of content, the cluster centroid is obtained, and then the Euclidean distance of the data packet to the cluster centroid is calculated. This distance characterizes the distribution status of packets. Ultimately, kernel density estimation curves~\cite{kamalov2020kernel} are employed to statistically analyze the distribution of Euclidean distances of the two types of content traffic, representing the distribution differences of traffic under content shift.

The distribution statistics of traffic for each environment shift are illustrated in Figure~\ref{diff}. Among them, \textit{none} represents the distribution of two groups of traffic without environment shift, and it can be observed that the data distribution almost overlaps in all categories of traffic. However, this phenomenon is nearly impossible to achieve in the real world. In all shift settings, the distributions of the two groups of data exhibit differences to varying degrees. Comparatively, content shift, application shift, resolution shift, and playback shift all induce significant data distribution differences. Consequently, these common environment shifts lead to different data distributions, further diminishing the accuracy of encrypted traffic classification.

\begin{figure*}[!t]
    \centering
    \includegraphics[width=0.95\textwidth]{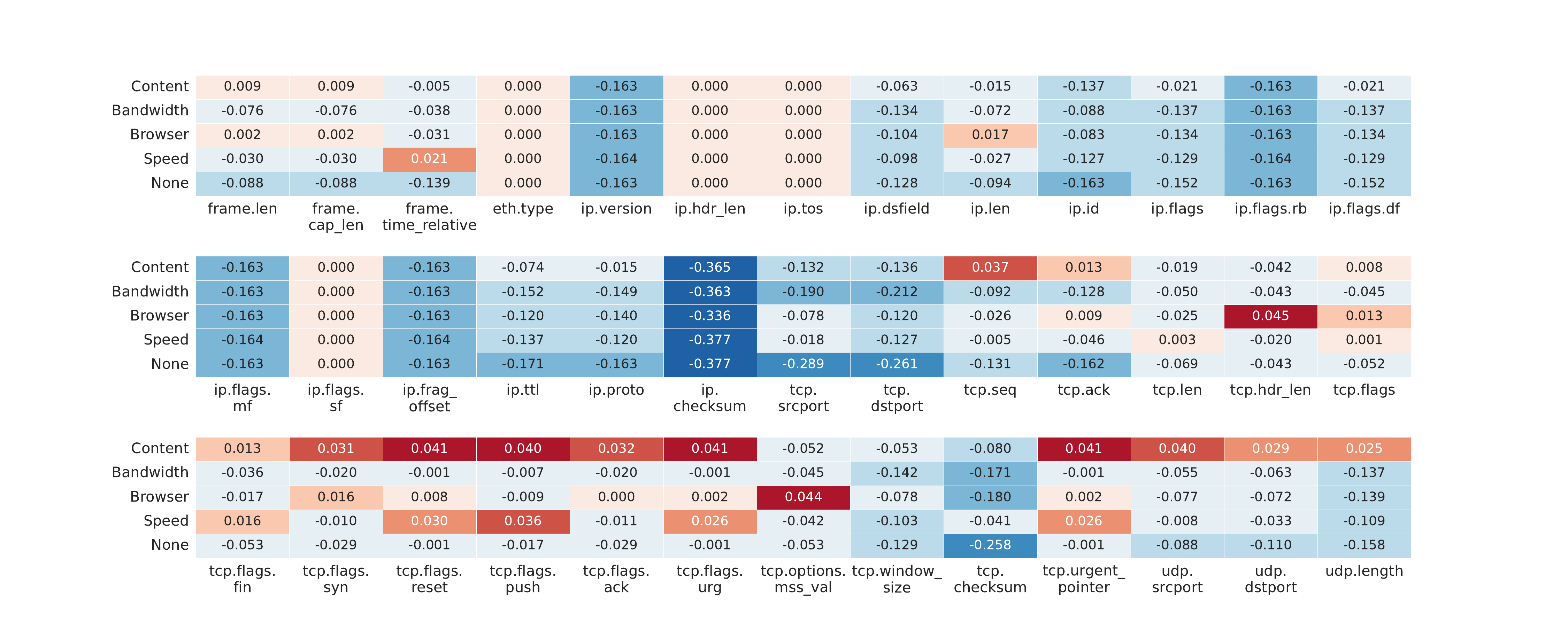}
    \caption{The results on JSD-based feature selection. The horizontal axis denotes the header fields and the vertical axis denotes specific environment shifts. The color gradient indicates the magnitude of the JSD difference, blue denotes the robust fields and red denotes the unstable fields, and the larger the difference, the stronger the color.}
    \label{JSD}
\end{figure*}

\subsubsection{Analysis on JSD-based Algorithm}

We examine the JSD-based algorithm's response to environment shifts in ``browsing'' traffic, focusing on content, bandwidth, browser, and speed shift factors. We assess feature robustness using inter-class and extra-class JSD comparisons, illustrated in Figure~\ref{JSD} with blue for stable (inter-class JSD < extra-class JSD) and red for unstable features (inter-class JSD > extra-class JSD). Without environment shifts, all features appear stable for browsing traffic identification. However, environment shifts introduce unstable features, notably affecting packet length and time.

We observe that IP header fields, such as \textit{IP DS field, IP id, IP flags, IP TTL}, remain robust across shifts, highlighting their significance for browsing traffic identification. Content shifts impact features the most, leaving only 20 robust fields, whereas bandwidth shifts have the least effect.

\subsubsection{Comparison with Baseline Feature Selection}

\begin{table}[!htb]
\setlength\tabcolsep{6pt}
\centering
\caption{Comparison results with baseline feature selection algorithms.}
\begin{threeparttable}
\begin{tabular}{c|c|c|c|c|>{\centering\arraybackslash}p{1.0cm}}
\hline
Algorithm &Pre&Rec&$F_{1}$&Acc&Dimension\\
\hline
Chi-Squared Test & 0.7179&	0.7390&	0.7056&	0.8107&	25\\
L1-LR&0.7216&	0.7478&	0.7247&	0.8511&	32\\
RFE&0.7251&	0.7562&	0.7219&	0.8441&	25\\
RFFI&0.7123	&0.7684	&0.7313&	0.8366&	25	\\
None & 0.7616 & 0.7210 & 0.7380 & 0.8406 & 39 \\
\textbf{JSD Algorithm} &	\textbf{0.8250}&	\textbf{0.8049}&	\textbf{0.8124}&	\textbf{0.8979}&	\textbf{25}\\
\hline
\end{tabular}
\label{feature_select}
\end{threeparttable}
\end{table}

The aim of the JSD feature selection algorithm is to evaluate the differences in feature distribution and select robust features under environment shifts. In this section, we compare the performance of the JSD feature selection algorithm with other feature selection algorithms as described below. (1) Chi-Squared Test~\cite{hakim2019influence}, which evaluates the dependence between features and target variables using chi-square statistics. (2) L1-Regularized Logistic Regression (L1-LR)~\cite{li2019tuning}, which applies L1 regularization to logistic regression models, inducing sparsity and performing feature selection. (3) Recursive Feature Elimination (RFE)~\cite{jeon2020hybrid}, which iteratively removes the weakest features while building a model to find the optimal feature subset. and (4) Random Forest Feature Importance (RFFI)~\cite{speiser2019comparison}, which ranks features based on their importance in an ensemble of decision trees. The Chi-Squared Test, RFE and RFFI aim to score and rank features. To clearly compare the performance of them with that of the JSD algorithm, we ensure that the dimensions of their respective features remain consistent with the dimensions of the JSD features. While the L1-LR indirectly achieves feature selection by reducing the coefficients of some features to zero, therefore, we maintain its original feature selection dimensions.

Comparison results, as shown in Table~\ref{feature_select}, indicate that none of these feature selection algorithms can outperform the JSD algorithm under environment shifts. Additionally, only L1-LR provides better classification results than those obtained without feature selection, whereas the other two feature selection algorithms provide lower accuracy. Consequently, the JSD algorithm outperforms state-of-the-art algorithms under environment shifts.

\subsection{Analysis on GraphSAT}

\begin{table}[!t]
\setlength\tabcolsep{5pt}
\centering
\caption{The results of analysis on GraphSAT and comparision with GCN~\cite{jiang2019semi}, GraphSAGE~\cite{DBLP:conf/nips/HamiltonYL17}, and GAT~\cite{velivckovic2017graph}.}
\begin{threeparttable}

\begin{tabular}{c|c|c|c|c|c|c}
\hline
Model&Pre&Rec&$F_{1}$&Acc&Time (ms)&Params\\
\hline
GCN&	0.8799&	0.8711&	0.8752&	0.9358&4.99& \textbf{38149}	\\
GraphSAGE&0.8998&	0.8953&	0.8973&	0.9478&\textbf{4.17}&	75269\\
GAT&0.8892&		0.8988&	0.8928&0.9462&	5.15&	38917\\
\textbf{GraphSAT} &	\textbf{0.9028}&\textbf{0.8956}&\textbf{0.8991}&	\textbf{0.9516}& 4.82&	43013\\
\hline
\end{tabular}
\label{model}

\end{threeparttable}

\end{table}
In this section, we conduct a comprehensive analysis of the GraphSAT model in relation to other traditional GNN models, specifically, GCN, GraphSAGE, and GAT. Our goal is to evaluate the efficacy and efficiency of our proposed model in the context of encrypted traffic application classification. The results present in Table~\ref{model} demonstrate that GraphSAT exhibits better performance across various evaluation metrics, including Pre, Rec, $F_{1}$, and Acc, highlighting its effectiveness in handling encrypted traffic classification tasks.

Furthermore, our analysis reveals that GraphSAT offers a competitive balance between accuracy and computational demands. The optimal combination of accuracy, time, and parameters count positions GraphSAT as a more favorable choice for encrypted traffic classification in comparison to the other GNN models.

\subsection{Analysis on Open-World Classification}
\begin{figure}
    \centering
    \includegraphics[width=\linewidth]{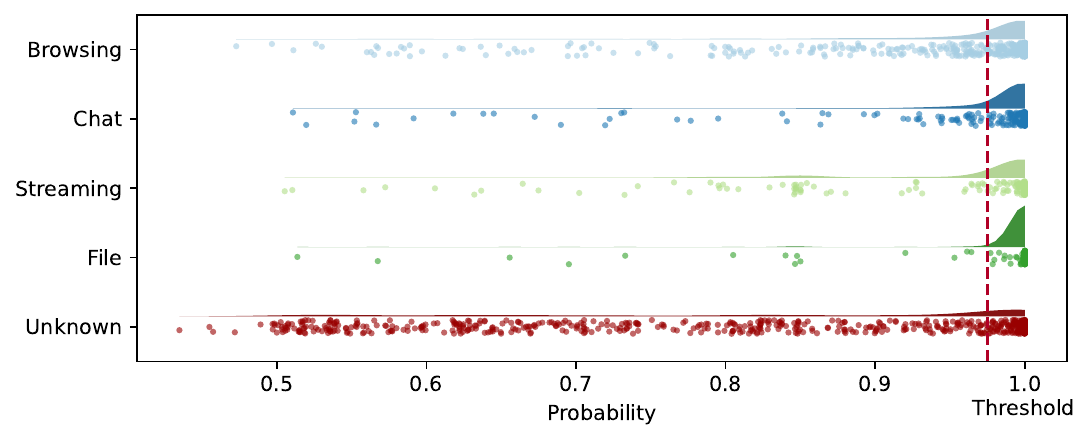}
    \caption{The results on open-world classification, where the email class is removed from the training set but kept in the test set as an unknown class. The results show that 70\% of the traffic falls below the threshold, while over 85\% of the known traffic is above the threshold.}
    \label{open-world}
\end{figure}

In the open-world, challenges arise not only from environment shifts but also from the presence of unknown class traffic, which is not seen during training. To investigate the performance of FG-SAT in such open-world scenarios, we incorporate an additional unknown class in the test set that is absent from the training set. The experimental results, as depicted in Figure~\ref{open-world}, demonstrate some noteworthy patterns.

We observe that the prediction probability for known classes is relatively high, with 85\% of them exceeding 0.975. In contrast, the prediction probability for 70\% of the unknown traffic falls below 0.975. This finding indicates that FG-SAT is generally adept at detecting known classes with high confidence. However, when confronted with unknown traffic, the feature patterns of unknown traffic often exhibit substantial dissimilarities compared to those of known classes, which consequently leads FG-SAT to struggle with predicting them as one of the known classes. This results in lower classification probabilities for unknown traffic.

Given these findings, we propose that in open-world scenarios, the identification of unknown traffic can be effectively achieved by setting an appropriate threshold for the classifier. This approach allows FG-SAT to discriminate between known and unknown classes based on their classification probabilities.

\subsection{Parameter Tuning}
\label{sec:param}
In this section, we introduce the selection of parameters, including (1) the maximum packet number, and (2) the dimension of the hidden channel.

\subsubsection{The Maximum Packet Number}
\label{sec:pn}
\begin{figure}[!t]
    \centering
    \includegraphics[width=0.9\linewidth]{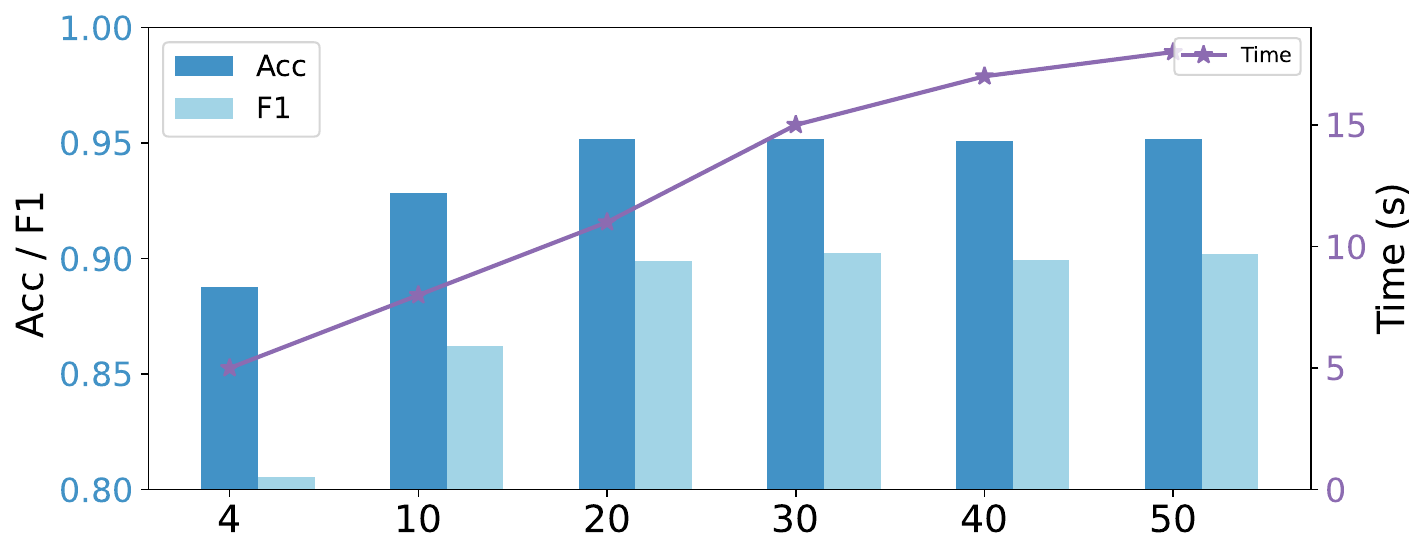}
    \caption{The results on the maximum packet number of FG-SAT.}
    \label{number}
\end{figure}

Flow duration can vary widely, impacting model efficiency. We adjust flow aggregation time by setting a maximum packet number, the results of which are illustrated in Figure~\ref{number}. Experiments with different packet counts show accuracy improves with higher counts up to a threshold. Analysis indicates accuracy gains level off beyond 20 packets, while aggregation time increases significantly. Therefore, we set the maximum packet count at 20.

\subsubsection{The Dimension of Hidden Channel}
\begin{figure}[!t]
    \centering
    \includegraphics[width=0.9\linewidth]{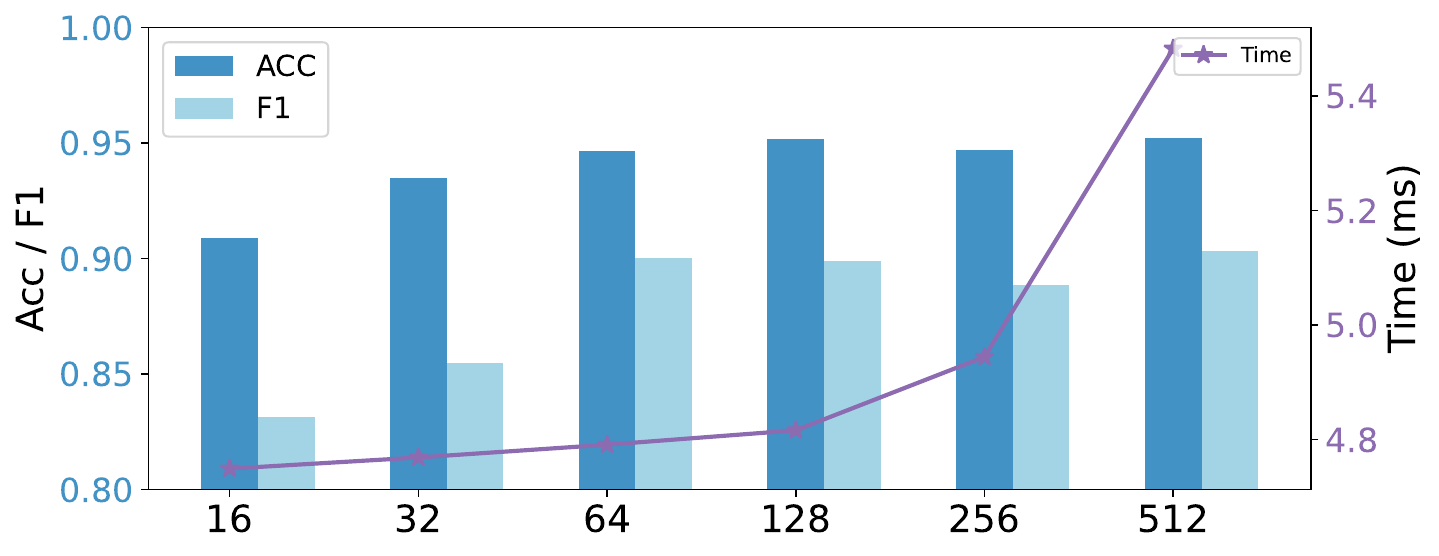}
    \caption{The results on the dimension of the hidden channel of FG-SAT.}
    \label{hidden}
\end{figure}

A larger hidden channel dimension enhances flow graph feature extraction and accuracy but increases computation time. Balancing accuracy and computational complexity is key. Through a series of experiments, as depicted in Figure~\ref{hidden}, we observe show that as the hidden channel dimension grows, accuracy improves up to a point. At a dimension of 128, accuracy plateaus, indicating an optimal balance between accuracy and processing time. Thus, we set the hidden channel dimension at 128 for efficient, effective encrypted traffic classification.

\section{Conclusion}
In this paper, we propose the Flow Graph for efficient encrypted traffic classification, which constructs flows as graphs to characterize the internal structure relationships and rich node attributes. To improve the generalization of the model to the real world, we propose a JSD-based feature selection algorithm that is able to select robust features under environment shifts. In addition, we design a fusion GraphSAGE and GAT classifier GraphSAT, which can efficiently and deeply learn Flow Graph features to achieve fast and accurate classification. We comprehensively evaluate the FG-SAT in three different scenarios of encrypted traffic classification tasks. It shows outstanding performance and outperforms state-of-the-art methods in terms of accuracy, time, parameters, and generalization ability. 

In our future work, we plan to improve the Flow Graph by investigating the influence of different edge relationships and assigning them varying weights. Additionally, we will further investigate the ability of FG-SAT to identify unknown categories. These efforts will enhance the effectiveness and robustness of our proposed method, as well as contribute to the advancement of the field of network traffic analysis.

\bibliographystyle{IEEEtran}
\bibliography{mybib}

\end{document}